\documentclass[twocolumn,showpacs,showkeys,superscriptaddress]{revtex4}
\usepackage{amsmath}
\usepackage{amssymb}
\usepackage{graphicx}

\newcommand{\bastar}{\begin{eqnarray*}}
\newcommand{\eastar}{\end{eqnarray*}}
\newskip\humongous \humongous=0pt plus 1000pt minus 1000pt

\relax
\newcommand{\bea}{\begin{eqnarray}}
\newcommand{\eea}{\end{eqnarray}}

\newcommand{\pro}{\partial}
\newcommand{\hn}{\hat n}
\newcommand{\pd}{\partial}
\newcommand{\mn}{{\mu \nu}}

\newcommand{\n}{\vec n}

\newcommand{\ba}{\begin{array}}
\newcommand{\ea}{\end{array}}
\newcommand{\Int}{{\displaystyle \int}}

\newcommand{\nn}{\nonumber}
\newcommand{\om}{\omega}
\begin{document}
\title{New Topological Structures of Skyrme Theory: Baryon 
Number and Monopole Number}
\author{Y. M. Cho}
\email{ymcho7@konkuk.ac.kr}
\affiliation{Institute of Modern Physics, Chinese Academy of
Science, Lanzhou 730000, China}
\affiliation{Administration Building 310-4,
Konkuk University, Seoul 143-701, Korea}
\affiliation{School of Physics and Astronomy, 
Seoul National University, Seoul 151-742, Korea}
\author{Kyoungtae Kimm}
\affiliation{Faculty of Liberal Education, 
Seoul National University, Seoul 151-747, Korea}
\author{J. H. Yoon}
\affiliation{Department of Physics, 
Konkuk University, Seoul 143-701, Korea}
\author{Pengming Zhang}
\affiliation{Institute of Modern Physics, Chinese Academy of
Science, Lanzhou 730000, China}

\begin{abstract}
~~~~~Based on the observation that the skyrmion in 
Skyrme theory can be viewed as a dressed monopole, 
we show that the skyrmions have two independent 
topology, the baryon topology $\pi_3(S^3)$ and 
the monopole topology $\pi_2(S^2)$. With this we propose 
to classify the skyrmions by two topological numbers 
$(m,n)$, the monopole number $m$ and the shell 
(radial) number $n$. In this scheme the popular 
(non spherically symmetric) skyrmions are classified 
as the $(m,1)$ skyrmions but the spherically symmetric 
skyrmions are classified as the $(1,n)$ skyrmions, 
and the baryon number $B$ is given by $B=mn$. 
Moreover, we show that the vacuum of the Skyrme theory has 
the structure of the vacuum of the Sine-Gordon theory 
and QCD combined together, which can also be classified 
by two topological numbers $(p,q)$. This puts the Skyrme 
theory in a totally new perspective. 

\end{abstract}
\maketitle

\section{Introduction}

The Skyrme theory has played an important role in physics. 
It has been proposed as an effective field theory of pion physics 
in strong interaction where the baryons appear as the skyrmions, 
topological solitons made of pions \cite{sky,witt}.  This view 
has been very successful, and the rich topological structure 
of the theory has advanced our understanding of the extended 
objects greatly \cite{rho,bogo,prep}. 

The construction of skyrmions as nuclei has a long history. 
A novel way to obtain non-spherically symmetric multi-skyrmions 
was developed based on the rational map, and the solutions have 
been associated with and compared to real nuclei \cite{bat,man}. 
And a systematic approach to construct the skyrmions with large 
baryon number numerically which have the shell strucutre has 
been developed \cite{houg,piet,sut}. This, with the improved 
computational power has made people construct skyrmions with 
the baryon number up to 108 \cite{sky108}. 

With the new development the Skyrme theory have had a 
remarkable progress recently. It has been able to provide a 
quantitative understanding of the spectrum of rotational 
excitations of carbon-12, including the excitation the Hoyle 
state which is essential for the generation of heavy nuclear 
elements in early universe \cite{hoyle,epel,lau}. And the 
spin-orbit interaction which is essential for the magic 
number of nuclei is investigated within the framework of 
Skyrme theory \cite{hal}. Moreover, a method to reduce 
the binding energy of skyrmions to a realistic level to 
improve the Skyrme model has been developed \cite{adam}. 
So by now in principle one could construct all nuclei as 
multi-baryon skyrmions and discuss the phenomenology
of nuclear physics, although the experimental confirmation 
of the theory is still in dispute. 

But the Skyrme theory has multiple faces. In addition 
to the well known skyrmions it has the (helical) baby 
skyrmion and the Faddeev-Niemi knot. Most importantly, 
it has the monopole which plays the fundamental 
role \cite{prl01,plb04,ijmpa08}. In this view all finite 
energy topological objects in the theory could be viewed 
either as dressed monopoles or as confined magnetic 
flux of the monopole-antimonopole pair. The skyrmion 
can be viewed as a dressed monopole, the baby skyrmion 
as a magnetic vortex created by the monopole-antimonopole 
pair infinitely separated apart, and the Faddeev-Niemi 
knot as a twisted magnetic vortex ring made of the helical 
baby skyrmion. This confirms that the theory can be 
interpreted as a theory of monopole in which the magnetic 
flux of the monopoles is confines and/or screened.

The fact that the skyrmion is closely related to the monopole 
has been appreciated for a long time. It has been well known 
that the skyrmions could actually be viewed as the monopoles 
regularized to have finite energy \cite{prl01,plb04,ijmpa08}. 
In fact it has been well appreciated that the rational map 
which plays the crucial role in the construction of 
the multi-skyrmions is exactly the $\pi_2(S^2)$ mapping 
which provides the monopole quantum number \cite{houg}. 
Nevertheless the skyrmions have always been classified 
by the baryon number given by $\pi_3(S^3)$, not by 
the monopole number $\pi_2(S^2)$. This was puzzling. 

The purpose of this paper is two-fold. We first show
that the skyrmions have two topology, the baryon topology 
and the monopole topology, so that they are classified 
by two topological numbers, the baryon number $B$ 
and the monopole number $M$. Moreover, we show 
that the baryon number can be replaced by the radial 
(shell) number which describes the $\pi_1(S^1)$ topology 
of radial excitation of multi-skyrmions. This is based on 
the observation that the SU(2) space $S^3$ has the Hopf 
fibering $S^3\simeq S^2\times S^1$ and that 
the Skyrme theory is described by two variables which 
naturally represent the $S^2$ and $S^1$ manifolds. 

Second, we show that the vacuum of the Skyrme 
theory has the structure of the vacuum of the Sine-Gordon 
theory and QCD combined together, which can also be 
classified by two topological numbers $(p,q)$.

The paper is organized as follows. In Section II we 
briefly review the old skyrmions for later purpose. 
In Section III we show that the skyrmions carry two 
topological numbers, the baryon number $b$ and 
the monopole number $m$. Moreover, we show that 
the baryon number can be replaced by the radial 
(shell) number $n$, so that they can be classified by 
$(m,n)$. In this scheme the baryon number is given 
by $b=mn$. In Section IV we discuss the vacuum 
structure of the Skyrme theory, and show that it has 
the structure of the vacuum of the Sine-Gordon theory 
combined with the vacuum of the SU(2) QCD. This 
tells that it can be classified by two topological numbers 
denoted by $(p,q)$, where $p$ and $q$ represent 
the $\pi_1(S^1)$ topology of the Sine-Gordon theory 
and the $\pi_3(S^2)$ topology of QCD vacuum. Finally 
in Section V we discuss the physical implications of 
our results.

\section{Skyrme Theory: A Review}

To see this let $\om$ and $\hn~(\hn^2=1)$ be the massless 
sigma field and the normalized pion field in Skyrme 
theory, and consider the Skyrme Lagrangian
\bea
&{\cal L}=\dfrac{\kappa^2}{4} {\rm tr}~L_\mu^2+\dfrac{\alpha}{32} 
{\rm tr} \left( \left[ L_\mu, L_\nu \right] \right)^2  \nn\\
&= - \dfrac{\kappa^2}{4} \Big[ \dfrac{1}{2} (\pd_\mu \om)^2
+2 \sin^2 \dfrac{\om}{2} (\pd_\mu \hn)^2 \Big]  \nn\\
&-\dfrac{\alpha}{8}  \sin^2 \dfrac{\om}{2}
\Big[(\pd_\mu \om)^2 (\pd_\nu \hn)^2-(\pd_\mu \om \pd_\nu \om)
(\pd_\mu \hn)\cdot (\pd_\nu \hn) \Big]   \nn\\
&+\dfrac{\alpha}{4} \sin^4 \dfrac{\om}{2} (\pd_\mu \hn 
\times \pd_\nu \hn)^2, \nn\\,
&L_\mu = U\pd_\mu U^{\dagger}, \nn\\
&U = \exp (\dfrac{\om}{2i} \vec \sigma \cdot \hn)
= \cos \dfrac{\om}{2} - i (\vec \sigma \cdot \hn)
\sin \dfrac{\om}{2},
\label{slag}
\eea
where $\kappa$ and $\alpha$ are the coupling constants.
Notice that $\hn$ and $\om$ naturally describe 
the $S^2$ and $S^1$ manifold. With
\bea
&U =\sigma- i \vec \sigma \cdot \vec \pi, \nn\\
&\sigma=\cos \dfrac{\omega}{2},
~~~\vec \pi= \hn \sin \dfrac{\omega}{2}, 
~~~(\sigma^2 + \vec \pi^2 = 1),
\label{sm}
\eea
the Lagrangian (\ref{slag}) has the familiar form
\bea
&{\cal L} = -\dfrac{\kappa^2}{2} \big((\partial_\mu \sigma)^2
+(\partial_\mu \vec \pi)^2 \big) \nn\\
&-\dfrac{\alpha}{4} \big((\partial_\mu \sigma \partial_\nu \vec \pi
- \partial_\nu \sigma \partial_\mu \vec \pi)^2
+ (\partial_\mu \vec \pi \times \partial_\nu \vec \pi)^2 \big) \nn\\
&+\dfrac{\lambda}{4} (\sigma^2 + \vec \pi^2 - 1),
\label{smlag}
\eea
where $\lambda$ is a Lagrange multiplier. In this form
$\sigma$ and $\vec \pi$ represent the sigma and 
pion fields, so that the Skyrme theory describes 
the pion physics.

The Lagrangian has a hidden $U(1)$ gauge symmetry
as well as a global $SU(2)_L \times SU(2)_R$ symmetry 
\cite{plb04,ijmpa08}. The global $SU(2)$ symmetry is 
obvious, but the hidden $U(1)$ symmetry is not. It comes 
from the fact that $\hn$ has an invariant subgroup $U(1)$. 
To see this, we reparametrize $\hn$ by the $CP^1$ field 
$\xi$,
\bea
\n = \xi^\dag \vec \sigma \xi,
~~~\xi^\dag \xi=1.
\label{ndef}
\eea
and find that under the $U(1)$ gauge transformation 
of $\xi$
\bea
\xi \rightarrow \exp (i\theta(x)) \xi,
\label{u1}
\eea
$\hn$ (and $\pro_\mu \hn$) remains invariant. 
Now, we introduce the composite gauge potential 
$B_\mu$ and the covariant derivative $D_\mu$ 
which transforms gauge covariantly under (\ref{u1}) 
by
\begin{gather}
B_\mu = -i \xi^\dagger \pd_\mu \xi,
~~~D_\mu \xi = (\pd_\mu -i B_\mu )\xi.
\end{gather}
With this we have the following identities,
\begin{gather}
(\pd_\mu \hn )^2 = 4 |D_\mu \xi |^2,  \nn\\
\pd_\mu \hn \times \pd_\nu \hn =-2i \Big[(\pd_\mu \xi^\dagger)
(\pd_\nu \xi) - (\pd_\mu \xi^\dagger)(\pd_\nu\xi)\Big] \hn \nn\\
=2 G_\mn \hn,~~~~~G_\mn=\pd_\mu B_\nu - \pd_\nu B_\mu.
\end{gather}
Furthermore, with the Fierz' identity
\begin{gather}
\sigma_{ij}^a \sigma_{kl}^a = 2\delta_{il}\delta_{jk} 
- \delta_{ij}\delta_{kl},
\end{gather}
we have
\begin{gather}
\pd_\mu \hn \cdot \pd_\nu \hn 
=2 \pd_\mu (\xi^\dagger_i \xi_j) 
\pd_\nu (\xi^\dagger_j \xi_i)  \nn\\
= 2\Big[ (\pd_\mu\xi^\dagger\xi )(\pd_\nu \xi^\dagger\xi)
+(\pd_\mu \xi^\dagger)( \pd_\nu \xi )  \nn\\
+(\partial_\nu \xi^\dagger)( \partial_\mu \xi) 
+(\xi^\dagger \pd_\mu \xi)(\xi^\dagger \pd_\nu \xi) \Big] \nn\\
= 2\Big[ (D_\mu \xi )^\dagger (D_\nu \xi) 
+ (D_\nu \xi )^\dagger (D_\mu \xi)\Big]. 
\end{gather}
From this we can express (\ref{slag}) by
\begin{gather}
{\cal L} = - \dfrac{\kappa^2}{4} 
\Big[ \dfrac{1}{2} (\pd_\mu \om)^2
+8 \sin^2 \dfrac{\om}{2} |D_\mu \xi|^2 \Big]  \nn\\
-\dfrac{\alpha}{2}  \sin^2 \dfrac{\om}{2}
\Big[(\pd_\mu \om)^2 |D_\mu \xi|^2
-(\pd_\mu \om \pd_\nu \om) 
(D_\mu \xi )^\dagger (D_\nu \xi)  \Big]  \nn\\
-\alpha \sin^4 \dfrac{\om}{2} F_\mn^2, 
\end{gather}
which is explicitly invariant under the $U(1)$ gauge 
transformation (\ref{u1}). So replacing $\hn$ by 
$\xi$ in the Lagrangian we can make the hidden 
$U(1)$ gauge symmetry explicit. In this form 
the Skyrme theory becomes a self-interacting 
$U(1)$ gauge theory of $CP^1$ field coupled to 
a massless scalar field.

From (\ref{slag}) we have the following equations of 
motion \cite{prl01,plb04,ijmpa08}
\bea
&\pd^2 \om -\sin\om (\pd_\mu \hn)^2
+\dfrac{\alpha}{8 \kappa^2} \sin\om (\pd_\mu \om
\pd_\nu \hn -\pd_\nu \om \pd_\mu \hn)^2  \nn\\
&+\dfrac{\alpha}{\kappa^2} \sin^2 \dfrac{\om}{2}
\pd_\mu \big[ (\pd_\mu \om \pd_\nu \hn
-\pd_\nu \om \pd_\mu \hn) \cdot \pd_\nu \hn \big] \nn\\
&- \dfrac{\alpha}{\kappa^2} \sin^2 \dfrac{\om}{2}
\sin\om (\pd_\mu \hn \times \pd_\nu \hn)^2 =0, \nn \\
&\pd_\mu \Big\{\sin^2 \dfrac{\om}{2}  \hn \times
\pd_\mu \hn + \dfrac{\alpha}{4\kappa^2} \sin^2 \dfrac{\om}{2}
\big[ (\pd_\nu \om)^2 \hn \times \pd_\mu \hn  \nn\\
&-(\pd_\mu \om \pd_\nu \om) \hn \times \pd_\nu \hn \big] \nn\\
&+\dfrac{\alpha}{\kappa^2} \sin^4 \dfrac{\om}{2} (\hn \cdot
\pd_\mu \hn \times \pd_\nu \hn) \pd_\nu \hn \Big\}=0.
\label{skeq1}
\eea
It has two interesting limits. First, in the spherically symmetric 
limit
\bea
\om = \om (r),~~~~~\hn = \pm \hat r,
\label{skans}
\eea
it is reduced to 
\bea
&\dfrac{d^2 \om}{dr^2} +\dfrac{2}{r} \dfrac{d\om}{dr}
-\dfrac{2\sin\om}{r^2} +\dfrac{2\alpha}{\kappa^2}
\Big[\dfrac{\sin^2 (\om/2)}{r^2} \dfrac{d^2 \om}{dr^2}  \nn\\
&+\dfrac{\sin\om}{4 r^2} (\dfrac{d\om}{dr})^2
-\dfrac{\sin\om \sin^2 (\om /2)}{r^4} \Big] =0.
\label{skeq2}
\eea
Adopting the spherically symmetric ansatz (\ref{skans}) 
imposing the boundary condition 
\bea
\om(0)= 2\pi,~~~~~\om(\infty)= 0,
\label{skbc}
\eea
we 
obtain the original skyrmion solution solving (\ref{skeq2}) which 
has the finite energy \cite{sky}. It carries the baryon number
\bea
&B =-\dfrac{1}{8\pi^2} \Int \epsilon_{ijk} \pd_i \om
\big[\hn \cdot (\pd_j \hn \times \pd_k \hn) \big]
\sin^2 \dfrac{\om}{2} d^3r \nn\\
&= 1,
\label{bn}
\eea
which represents the non-trivial homotopy $\pi_3(S^3)$
defined by $U$ in (\ref{slag}). 

Second, when  
\bea
\om= (2n+1) \pi,
\label{fadd}
\eea
the equation is reduced to
\bea
&\hn \times \pd^2 \hn+ \dfrac{\alpha}{\kappa^2} (\pd_\mu H_\mn)
\pd_\nu \hn = 0, \nn\\
&H_\mn= \hn \cdot (\pd_\mu \hn \times \pd_\nu \hn)
= \pd_\mu C_\nu - \pd_\nu C_\mu,
\label{sfeq}
\eea
where $C_\mu$ is the magnetic potential of $H_\mn$. 
This is the central equation of Skyrme theory which allows 
the monopole, the baby skyrmion, the twisted magnetic 
vortex, and the knot \cite{prl01,plb04,ijmpa08}. In fact
(\ref{sfeq}) has the monopole solution \cite{prl01,plb04,ijmpa08}
\bea
\hn = \pm \hat r.
\label{mono}
\eea
which carries the magnetic charge
\bea
M = \dfrac{\pm 1}{8\pi} \int \epsilon_{ijk} \big[\hat r
\cdot (\pd_i \hat r \times \pd_j \hat r)\big] d\sigma_k 
= \pm 1,
\eea
which represents the homotopy $\pi_2(S^2)$ defined 
by $\hn$. 

Notice that, with (\ref{fadd}) the Skyrme Lagrangian becomes 
the Skyrme-Faddeev Lagrangian,
\bea
{\cal L} \rightarrow -\dfrac{\kappa^2}{2} (\partial_\mu \hat
n)^2-\dfrac{\alpha}{4}(\partial_\mu \hat n \times
\partial_\nu \hat n)^2,
\label{sflag}
\eea
whose equation of motion is given by (\ref{sfeq}). This 
tells that the Skyrme-Faddeev theory is an essential 
ingredient, the back bone, of the Skyrme theory. 
As importantly, this reveals the ``missing link" between 
Skyrme theory and QCD. This is because we can derive 
the Skyrme-Faddeev Lagrangian directly from QCD, 
which shows that the two theories are related by
the Skyrme-Faddeev Lagrangian as the common 
denominator \cite{prl01,plb04,ijmpa08}.

\begin{figure}
\begin{center}
\includegraphics[height=5cm, width=7cm]{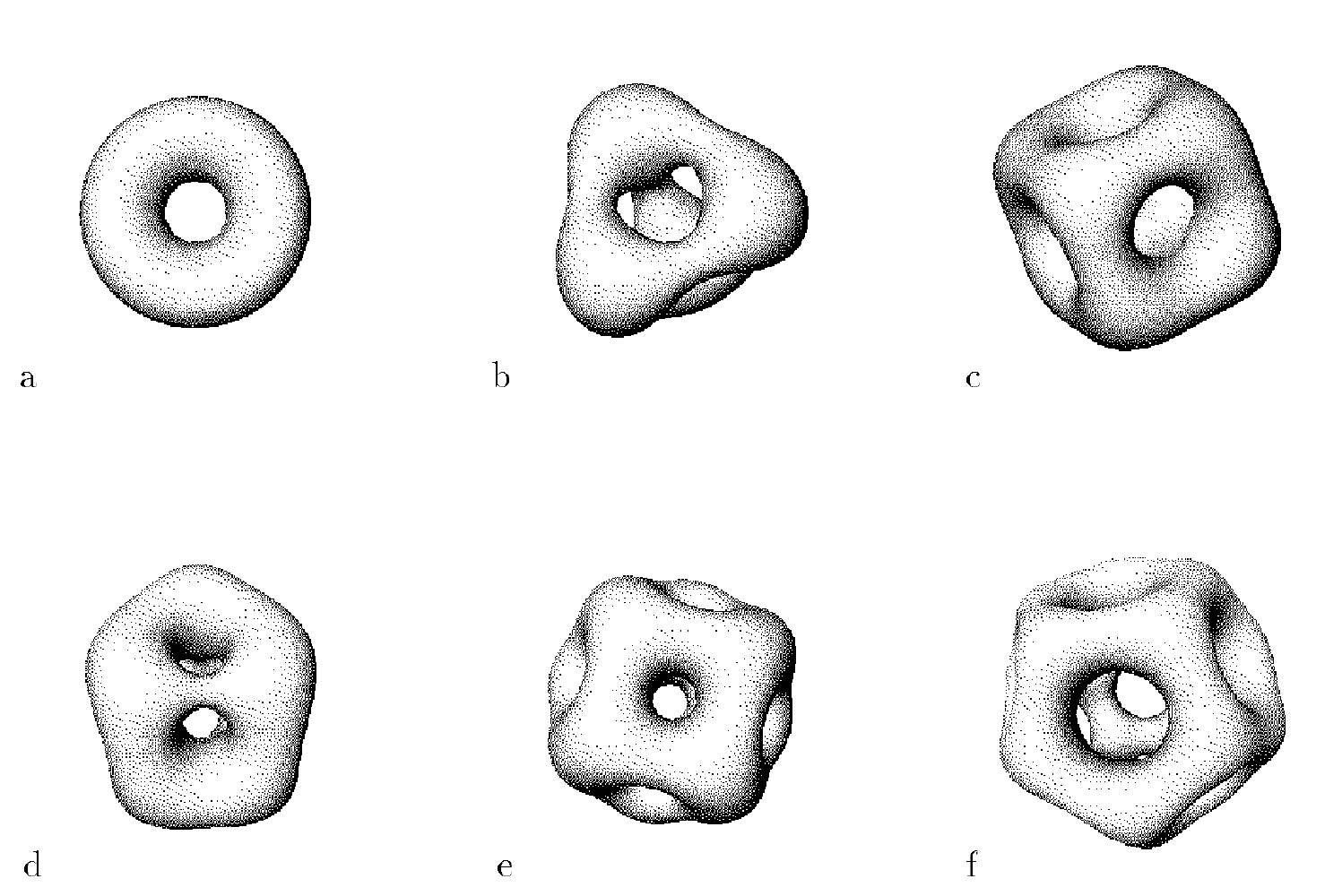}
\caption{\label{oldsol} The well known (non spherically 
symmetric) numerical multi-skyrmion solutions with 
baryon number 2,3,4,5,6, and 7, copied from 
Ref. \cite{houg}.}
\end{center}
\end{figure}

Solving (\ref{skeq1}) for multi-skyrmions numerically 
with (\ref{skbc}) choosing $\hn$ to describe an arbitrary 
rational map $\pi_2(S^2)$, one can obtain the well 
known (non spherically symmetric) multi-skyrmion 
solutions whose baryon number is given by the rational 
map number of $\hn$ numerically \cite{man,bat,houg,piet}. 
Some of these solutions are copied from Ref. \cite{houg} 
in Fig. \ref{oldsol}.  

In addition to these popular solutions we have other 
spherically symmetric multi-skyrmions. To obtain them 
notice that, although the SU(2) matrix $U$ is periodic 
in $\om$ variable by $4\pi$, $\om$ itself can take any 
value from $-\infty$ to $+\infty$. So we can obtain 
the spherically symmetric solution of (\ref{skeq2}) with 
the boundary condition
\bea
\om(0)= 2\pi n,~~~~~\om(\infty)= 0,
\label{skbcn}
\eea
with an arbitrary integer $n$ \cite{sky,witt,rho,bogo,prep}. 
Clearly they have the baryon number 
\bea
&B= \dfrac{1}{8\pi^2} \int \epsilon_{ijk} \pd_i \om
\big[\hat r \cdot (\pd_j \hat r \times \pd_k \hat r) \big]
\sin^2 \dfrac{\om}{2} d^3r \nn\\
&=\dfrac{1}{\pi} \int \sin^2 \dfrac{\om}{2} d\om=n.
\label{bnn}
\eea 
This means that the baryon number is given by the winding 
number $\pi_1(S^1)$ of $\om$, which is determined by 
the boundary condition (\ref{skbcn}). In Fig. \ref{skysol} 
we present the spherically symmetric skyrmions for 
$n=1,2,3,4,5,6,7$. 

\section{Baryon number versus Monopole number}

\begin{figure}
\begin{center}
\includegraphics[height=4.5cm, width=7cm]{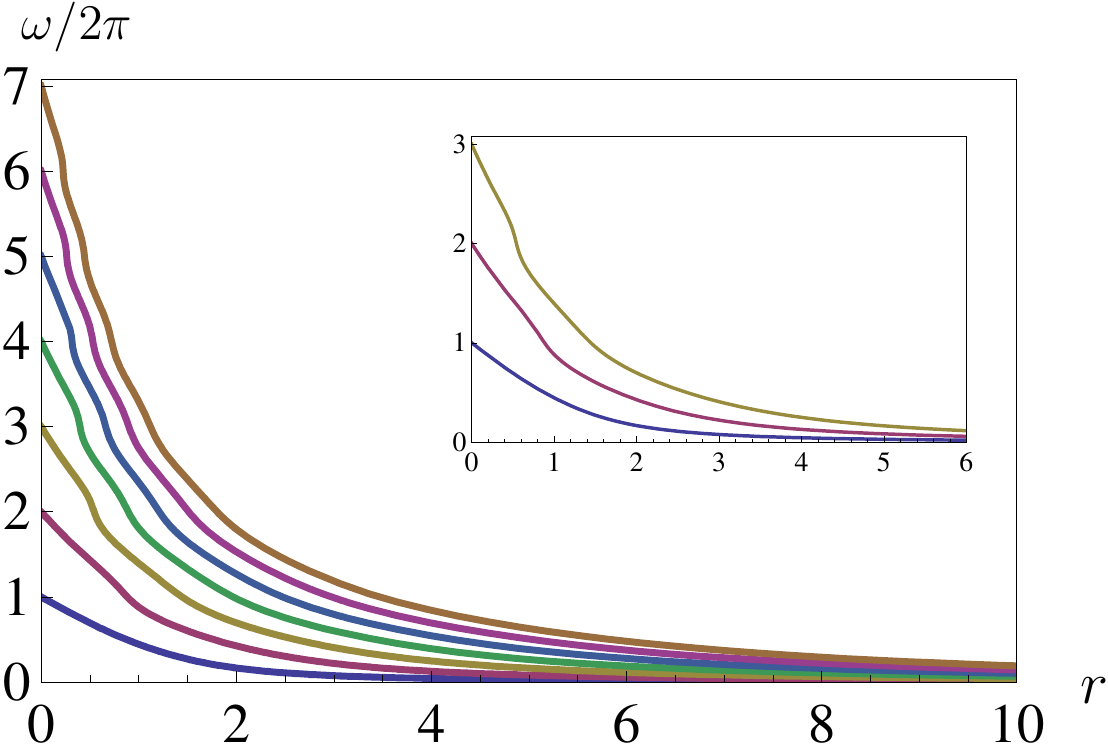}
\caption{\label{skysol} The spherically symmetric solutions
with baryon number 1,2,3,4,5,6,7, which should be 
contrasted with the popular multi-skyrmion solutions
shown in Fig. \ref{oldsol}.}
\end{center}
\end{figure}

The contrast between the two sets of solutions 
shown in Fig. \ref{oldsol} and Fig. \ref{skysol} is 
unmistakable, but this is not just in the appearence. 
They have a fundamental difference. Clearly the baryon 
number of the skyrmions shown in Fig. \ref{oldsol} 
is given by the rational map $\pi_2(S^2)$ defined 
by $\hn$ \cite{man,bat,houg,piet,sut}. Moreover, 
the rational map $\pi_2(S^2)$ of $\hn$ in the non 
spherically symmetric solutions is precisely the monopole 
topology of the skyrmion which determines the monopole 
number $M$ \cite{prl01,plb04,ijmpa08} 
\bea
M=\dfrac{1}{8\pi} \Int \epsilon_{ijk}\big[\hn\cdot
(\pd_i \hn  \times \pd_j \hn) \big] d\sigma_k=m.
\label{mono}
\eea 
And clearly this monopole number is different from 
the baryon number given by the $\pi_3(S^3)$ topology.

However, the baryon number of the skyrmions shown 
in Fig. \ref{skysol} is given by the winding number 
$\pi_1(S^1)$ of $\om$ which has nothing to do with 
the rational map of $\hn$. Moreover, the monopole 
number of these solutions given by the rational map 
$\pi_2(S^2)$ of $\hn$ is
\bea
M=\dfrac{1}{8\pi} \Int \epsilon_{ijk}\big[\hat r \cdot
(\pd_i \hat r  \times \pd_j \hat r) \big] d\sigma_k=1.
\label{mono1}
\eea 
This tells that the baryon number and the monopole 
number of these solutions are different. 

This shows that the skyrmions actually have {\it two} 
topological numbers $B$ and $M$ which are in principle 
different. But this was not evident in the popular skyrmion 
solutions because they have $B=M$. But obviously 
the spherically symmetric solutions have two topological 
numbers, the baryon number $B=n$ and the monopole 
number $M=1$ \cite{sky,man}. This proves that 
the skyrmions do have two topology denoted by $(b,m)$, 
the $\pi_3(S^3)$ which describes the baryon number 
$b$ and the $\pi_2(S^2)$ which describes the monopole 
number $m$. But so far this important point has been 
completely neglected. 

Moreover, the integer $n$ in (\ref{skbcn}) has another 
meaning. It describes the $\pi_1(S^1)$ topology of 
the angular variable $\om$ which depends only on 
the radial coordinate $r$. Moreover, it could be viewed 
as the radial, or more properly the shell quantum number, 
since the spherically symmetric solutions can be viewed 
as the generalization of the original skyrmion which 
has radially excited shells where $n$ describes the number 
of the shells. This implies that we could also classify 
the skyrmions by $(m,n)$, the $\pi_2(S^2)$ topology 
of $\hn$ and $\pi_1(S^1)$ topology of $\om$. 

In this scheme the well known (non spherically symmetric)
numerical multi-skyrmion solutions shown in the earlier 
works \cite{bat,houg} become the $(m,1)$ skyrmions, 
but those shown in Fig. \ref{skysol} become the $(1,n)$ 
skyrmions. This strongly implies that the baryon number 
is made of two parts, the $\pi_2(S^2)$ of $\hn$ and 
$\pi_1(S^1)$ of $\om$. 

To amplify this point we notice the followings. First, 
the $S^3$ space (both the real space and the target 
space) in $\pi_3(S^3)$ admits the Hopf fibering 
$S^3\simeq S^2\times S^1$. Second, the two variables 
$\hn$ and $\om$ of the Skyrme theory naturally represent 
$S^2$ and $S^1$. So the baryon number of the $(m,n)$ 
skyrmion is given by
\bea
&B =-\dfrac{1}{8\pi^2} \Int \epsilon_{ijk} \pd_i \om
\big[\hn \cdot (\pd_j \hn \times \pd_k \hn) \big]
\sin^2 \dfrac{\om}{2} d^3r \nn\\
&=-\dfrac{1}{8\pi^2} \Int \pd_i \om
\big[\hn \cdot (\pd_j \hn \times \pd_k \hn) \big]
\sin^2 \dfrac{\om}{2} dx^i \wedge dx^j \wedge dx^k \nn\\
&=-\dfrac{1}{8\pi^2} \Int \sin^2 \dfrac{\om}{2} d\om 
\times \epsilon_{ijk} \big[\hn \cdot (\pd_i \hn 
\times \pd_j \hn) \big] d\Sigma_k \nn\\
&=\dfrac{n}{8\pi} \Int \epsilon_{ijk} \big[\hn \cdot 
(\pd_i \hn \times \pd_j \hn) \big] d\Sigma_k=mn,
\label{bmn}
\eea
where $d\Sigma_k=\epsilon_{ijk} dx^i \wedge dx^j/2$.
Clearly the last integral is topologically equivalent 
to (\ref{mono}), which assures the last equality. 
This shows that the baryon number of the skyrmion 
can be decomposed to the monopole number and the shell 
number. Obviously both $(m,1)$ and $(1,n)$ skyrmions 
are the particular examples of this.  

At this point it must be emphasized that the shell 
number $n$ has first been introduced by Manton and 
Piette \cite{piet}. They have noticed that the skyrmions 
can be generalized to have the multiple shell structure 
which can be expressed by the shell number. Moreover, 
they have shown that this shell structure is very 
useful to construct the multi-skyrmions which have 
a large baryon number even for the non spherically 
symmetric skyrmions. 

What we propose here is that the baryon number can 
be decomposed to the monopole number and the shell 
number, and that this shell number could be interpreted 
to represent an independent topology of the skyrmion. 
This follows from the fact that the skyrmions have 
two independent topology, the monopole topology 
$\pi_2(S^2)$ and the baryon topology $\pi_3(S^3)$. 
Given this fact, the natural question is how they 
are related. The answer is that the baryon topology 
is made of the monopole topology and the shell topology, 
and the baryon number is given by the product of 
the monopole number and the shell number.  

\begin{figure}
\begin{center}
\includegraphics[height=4.5cm, width=7cm]{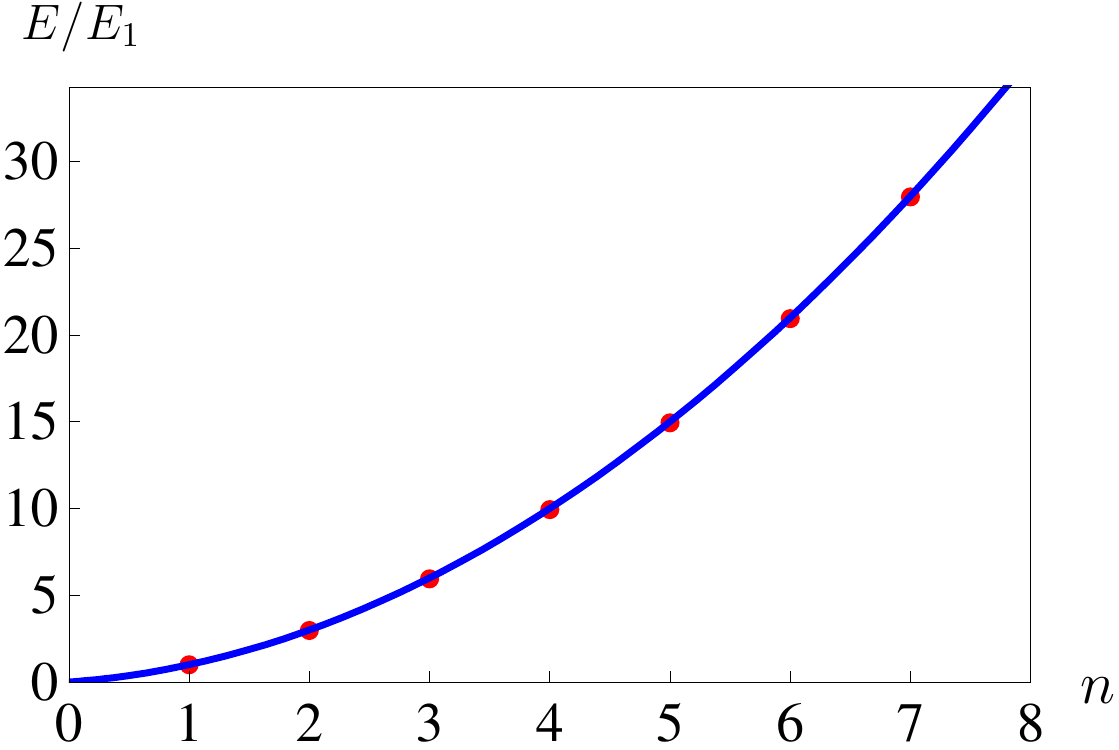}
\caption{\label{En} The energy of the spherically symmetric 
solutions with baryon number 1,2,3,4,5,6,7. The numerical fit 
(the blue curve) and the $n(n+1) E_1/2$ curve (the green 
curve) are almost indistinguishable.}
\end{center}
\end{figure}

This is based on two facts. First, the Skyrme theory 
is made of two variables, the $S^2$ variable $\hn$ 
which represents the $\pi_2(S^2)$ topology and 
the $S^1$ variable $\om$ which represents the shell 
topology $\pi_1(S^1)$. Second, the baryon topology 
is described by both $\om$ and $\hn$, but the monopole 
topology is described only by $\hn$. So it becomes 
only natural that $\om$ changes the monopole topology 
to the baryon topology, adding the shell structure 
to the monopole topology. It is this separation of 
the roles of the two variables which allows us to replace 
the baryon topology with the shell topology in Skyrme 
theory. Obviously this is best demonstrated in 
the spherically symmetric skyrmions.  

An interesting feature of the spherically symmetric 
solutions is that whenever the curve passes through 
the values $\om=2\pi n$, it become a bit steeper. There 
is a good reason why this is so. As we will see these 
points are the vacua of the theory, and the steep slopes 
shows that the energy likes to be concentrated around 
these vacua. So these steep slopes are not an irregularity, 
but just what is expected.

\begin{figure}
\begin{center}
\includegraphics[height=4.5cm, width=7cm]{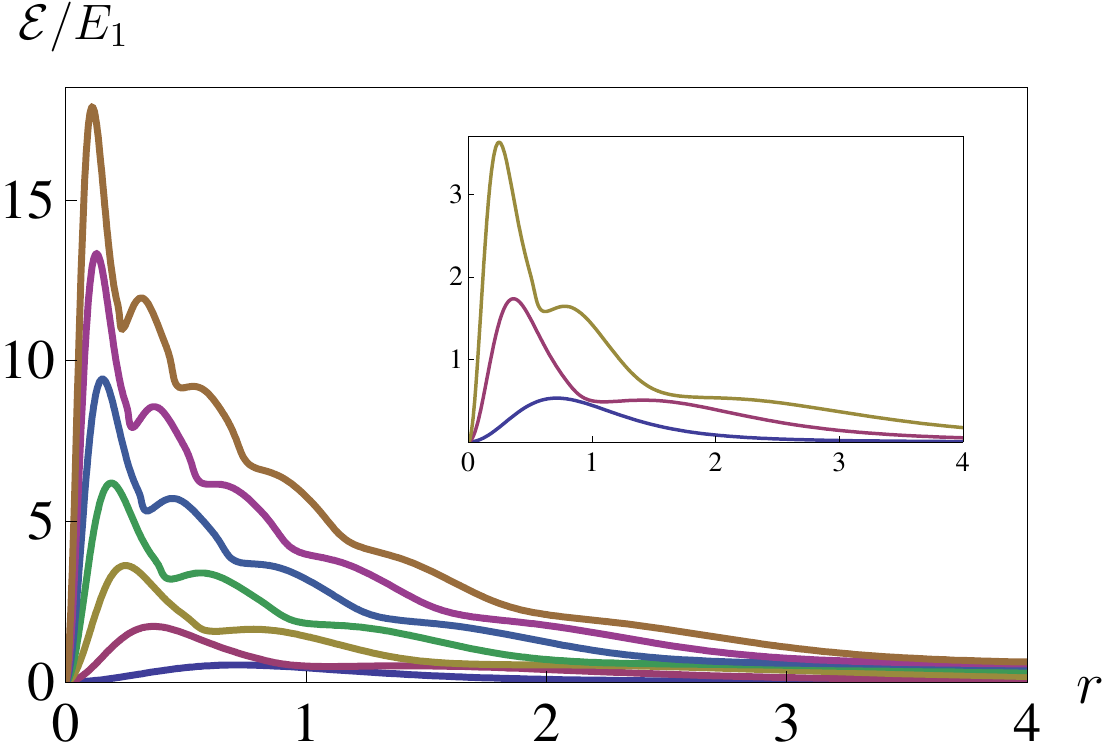}
\caption{\label{eden} The 
energy density of the spherically 
symmetric skyrmions with baryon number 1,2,3,4,5,6,7.}
\end{center}
\end{figure}

We can easily calculate the energy of the spherically 
symmetric solutions from \cite{ijmpa08}
\bea
&E =\dfrac{\pi \kappa^2}{2}  \Int_0^\infty \Big\{\Big(r^2
+\dfrac{2\alpha}{\kappa^2} \sin^2\dfrac{\om}{2}\Big)
\Big(\dfrac{d\om}{dr}\Big)^2  \nn\\
&+8 \Big(1+\dfrac{\alpha}{2\kappa^2 r^2} \sin^2 
\dfrac{\om}{2} \Big) \sin^2 \dfrac{\om}{2} \Big\} dr  \nn\\ 
&=\pi {\sqrt \alpha} \kappa \Int^{\infty}_{0}
\Big[x^2 \left(\dfrac{d\om}{dx}\right)^2
+ 8 \sin^2{\dfrac{\om}{2}} \Big] dx,
\label{sken}
\eea
where $x=(\kappa/\sqrt{\alpha})~r$ is a dimensionless 
variable. The result is shown in Fig. \ref{En}. Numerically 
the baryon number dependence of the energy is given 
by \cite{rho,bogo,prep}
\bea
E_n\simeq \dfrac{n(n+1)}{2} E_1.
\label{en}
\eea
This could be understood as follows. Roughly speaking, 
the kinetic energy (first part) and the potential energy 
(second part) of (\ref{sken}) become proportional to $n^2$ 
and $n$, and the two terms have equal contribution due to 
the equipartition of energy. But the truth is more complicated 
than this, and we need a mathematical explanation of this. 

The energy density of the solutions is shown in Fig. \ref{eden}. 
The $B=n$ solution has $n$ local maxima, which tells that it is 
made of $n$ unit skyrmions which make spherical shells. 
Moreover, as we have remarked the shells are located at 
the vacuum points $\om=2\pi n$.

Of course, these spherically symmetric skyrmions are 
precisely the multi-skyrmions that Skyrme originally 
proposed as nuclei which have baryon number larger 
than one \cite{sky,witt,rho,bogo,prep}. But they become 
unstable and can decay to the lower energy skyrmions, 
because the energy $E_n$ gets bigger than the $n$ sum 
of $E_1$. This is not so for the popular (non spherically 
symmetric) multi-skyrmions which have positive binding 
energy. Because of this the spherically symmetric skyrmions 
have been dismissed as uninteresting. 

But our analysis makes them more interesting. First of all, 
they demonstrate that skyrmions actually have two topological 
numbers, the baryon number and the monopole number, which 
are different. Moreover, they show that the skyrmions can 
be made to have the shell structure. As importantly, they 
tell that the shell number, together with the monopole 
number, determines the baryon number. 

Clearly the shell structure can also be implemented to 
the $(m,1)$ skyrmions shown in Fig. \ref{oldsol}. To see 
this we generalize the boundary condition (\ref{skbcn}), 
keeping the rational map number $m$ of $\hn$ the same 
but requiring \cite{piet}
\bea
&\om(r_k)=2\pi k,~~~(k=0,1,2,...n),  \nn\\
&r_0=0 ~\langle~r_1~\langle~...~\langle~r_n=\infty.
\eea
With this we could find the $(m,n)$ skyrmion numerically 
minimizing the energy varying $r_k~(k=1,2,...,n-1)$. This 
way we can add the shell structure and the shell number 
to the $(m,1)$ skyrmion. 
 
Now, one might ask about the stability of the $(m,n)$ skyrmions.
In general they may not be stable. For example, the quadratic 
dependence of the topological number $n$ of the energy 
(\ref{en}) makes $(1,n)$ skyrmions energetically unstable. 
On the other hand, even when they decay, the topology of 
the solution must not change. In other words the baryon 
number and the monopole number must be conserved. This, 
together with $B=mn$, tells that the shell number should 
also be conserved. From this we conclude that, when an $(m,n)$ 
skyrmion decays to $(m_1,n_1)$ and $(m_2,n_2)$ skyrmions, 
we must have $n=n_1+n_2$ and $m=m_1+m_2$. This, of course, 
is what is expected.
 
The above discussion raises another deep question. As 
we have remarked, when $\om=(2n+1)\pi$  the Skyrme 
theory has knot solutions described by $\hn$ whose 
topology is given by $\pi_3(S^2)$, in addition to 
the skyrmion solutions \cite{prl01,plb04,ijmpa08}. If so, 
can we dress the knots with $\om$ to provide a new type of 
shell structure, and extend the knots to have two quantum 
numbers $\pi_1(S^1)$ and $\pi_3(S^2)$?  This is a mind 
boggling question.

\section{Multiple Vacua of Skyrme Theory}

Now we show that the Skyrme theory in fact has another 
very important topological structure, the topologically 
different multiple vacua. To see this, notice that 
(\ref{skeq1}) has the solution 
\bea
\om=2\pi p,~~~(p;~integer),
\label{skyvac}
\eea
independent of $\hn$. And obviously this is the vacuum 
solution. 

This tells that the Skyrme theory has multiple vacua 
classified by the integer $p$ which is similar to 
the Sine-Gordon theory. But unlike the Sine-Gordon 
theory, here we have the multiple vacua without any 
potential. Moreover, the above discussion tells that 
the spherically symmetric skyrmions connect and occupy 
the $p+1$ adjacent vacua. This means that we can 
connect all vacua with the spherically symmetric skyrmions. 
Of course, one could introduce such vacua in Skyrme theory 
introducing a potential term in the Lagrangian \cite{nitta}. 
This is not what we are doing here. We have these vacua 
without any potential. 

But this is not the end of the story. To see this notice that
(\ref{skyvac}) becomes the vacuum independent of $\hn$.
This means that $\hn$ can add the $\pi_3(S^2)$ topology 
to each of the multiple vacua classified by another integer 
$q$, because it is completely arbitrary. And this is precisely 
the knot topology of the QCD vacuum \cite{plb07}. 

This is not surprising. Given the fact that there is a deep 
connection between Skyrme theory and QCD, it is natural 
that the Skyrme theory and QCD have similar vacuum 
structure. To amplify this point, notice that the most 
general SU(2) QCD vacuum can be expressed in terms of 
a right-handed $SU(2)$ basis $(\hn_1,\hn_2,\hn_3=\hn)$ 
by \cite{plb07}
\bea
&\hat \Omega_\mu
=-\dfrac12 \epsilon_{ijk} (\hn_i\cdot \pd_\mu \hn_j)
~\hn_k.
\label{qcdvac}
\eea
Clearly it has the $\pi_3(S^3)$ topology of the mapping 
from the compactified 3-dimensional space to the SU(2) 
group space defined by $(\hn_1,\hn_2,\hn)$. But since 
$(\hn_1,\hn_2,\hn)$ is completely determined by $\hn$ 
up to the U(1) rotation which leaves $\hn$ invariant, 
(\ref{qcdvac}) also has the knot topology $\pi_3(S^2)$ 
which describes the mapping from the real space $S^3$ 
to the coset space $S^2$ of $SU(2)/U(1)$.

Now it must be clear why the vacuum of Skyrme theory 
has the same knot topology. As we have noticed, 
the Skyrme theory has the vacuum (\ref{skyvac}) 
independent of $\hn$, and this $\hn$ adds the knot 
topology $\pi_3(S^2)$ to the vacuum. Of course, 
in the Skyrme theory we do not need the vacuum 
potential (\ref{qcdvac}) to describe the vacuum. We 
only need $\hn$ which describes the knot topology. 
 
This confirms that the vacuum in Skyrme theory has 
the topology of the Sine-Gordon theory and QCD 
combined together. This means that the vacuum of 
the Skyrme theory can also be classified by two quantum 
numbers $(p,q)$, the $\pi_1(S^1)$ of $\om$ and 
$\pi_3(S^2)$ of $\hn$. And this is so without any 
extra potential. As far as we know, there is no other 
theory which has this type of vacuum topology. 

At this point we emphasize the followings. First, the knot 
topology of $\hn$ is different from the monopole topology 
of $\hn$. The monopole topology $\pi_2(S^2)$ is associated 
to the isolated singularities of $\hn$, but the knot topology 
$\pi_3(S^2)$ does not require any singularity for $\hn$. 
And for a classical vacuum $\hn$ must be completely 
regular everywhere. So only the knot topology, not the 
monopole topology, can not describe a classical vacuum. 
And this is precisely the vacuum topology of QCD.

Second, the knot topology of the vacuum is different 
from the Faddeev-Niemi knot that we have in the Skyrme 
theory \cite{prl01}. The Faddeev-Niemi knot is a unique
and real (i.e., physical) knot which carries energy, which 
is given by the solution of (\ref{sfeq}). In particular, 
we have the knot solution when $\om=(2n+1)\pi$. On the other 
hand, we have the knot of the vacuum when $\om= 2\pi p$. 
Moreover, the vacuum knot has no energy, and is not unique. 
While the Faddeev-Niemi knot is unique, there are infinitely 
many $\hn$ which describes the same vacuum knot topology. 
So obviously they are different. What is really remarkable 
is that the same $\hn$ has multiple roles. It describes 
the monopole topology, the knot topology of Faddeev-Niemi 
knot, and the knot topology of the vacuum.   

\section{Discussions}

Skyrme theory has been known to have rich topological 
structures. It has the Wu-Yang type monopoles, the skyrmions 
as dressed monopoles, the baby skyrmions and twisted 
magnetic flux, and the Faddeev-Niemi knots made of twisted 
magnetic vortex ring \cite{prl01,plb04,ijmpa08}. This 
makes the theory very important not only in high energy 
physics but also in condensed matter physics, in particular 
in two-gap superconductor and two-component Bose-Einstein 
condensates  \cite{plb04,ijmpa08,ruo,pra05}. 

Our analysis tells that the theory has more topology. 
In this paper we have shown that the skyrmions are not 
just the dressed monopoles but actually carry the monopole 
number, so that they can be classified by two topological 
numbers, the baryon number and the monopole number. 
Moreover, we have shown that here the baryon number could 
be replaced by the radial (shell) number, so that the skyrmions 
can be classified by two topological numbers $(m,n)$, 
the monopole number $m$ which describes the $\pi_2(S^2)$ 
topology of the $\hn$ field and the radial (shell) number 
$n$ which describes the $\pi_1(S^1)$ topology of the $\om$ 
field. In this scheme the baryon number $B$ is given by 
the product of two integers $B=mn$. This comes from 
the following facts. First, the SU(2) space $S^3$ admits 
the Hopf fibering $S^3\simeq S^2\times S^1$. Second, 
the Skyrme theory has two variables, the angular variable 
$\om$ which can represent the $\pi_1(S^1)$ topology and 
the coset variable $\hn$ which represents the $\pi_2(S^2)$ 
topology. 

In this view the popular (non spherically symmetric) 
skyrmions are classified as the $(m,1)$ skyrmions, and 
the radially excited spherically symmetric skyrmions 
are classified as the $(1,n)$ skyrmions. and we can
construct the $(m,n)$ skyrmions adding the shell structure 
to the $(m,1)$ skyrmions. Moreover, we have shown that 
the skyrmions, when they are generalized to have two 
topological numbers, should have the topological stability 
of the two topology independently. This is remarkable.

As importantly, we have shown that the Skyrme theory 
has multiple vacua. The vacuum of the theory has 
the structure of the vacuum of the Sine-Gordon theory 
and at the same time the structure of QCD vacuum. 
So the vacuum  can also be classified by two topological 
numbers $p$ and $q$ which represent the $\pi_1(S^1)$ 
topology of the $\om$ field and the $\pi_3(S^2)$ 
topology of the $\hn$ field. 

The fact that the vacuum of the Skyrme theory has 
the $\pi_1(S^1)$ topology is not surprising, considering
that it has the angular variable $\om$. Moreover, 
the fact that the vacuum of the Skyrme theory has 
the $\pi_3(S^2)$ topology of the QCD vacuum could 
easily be understood once we understand that the Skyrme 
theory is closely related to QCD. What is really remarkable 
is that it has both $\pi_1(S^1)$ and $\pi_3(S^2)$ topology
at the same time. As far as we understand there is no 
other theory which has this feature. This again is closely 
related to the fact that $S^3$ admits the Hofp fibering 
and that the theory has two variables $\om$ and $\hn$. 

This raises interesting questions. Can we generalize 
the Faddeev-Niemi knot to have the $\pi_1(S^1)$ 
topology? If so, how do we obtain such knot? Do we 
have the vacuum tunneling in Skyrme theory? What 
instanton can we have in this theory? 

Clearly the above observations put the Skyrme theory in 
a totally new perspective. Our results in this paper show 
that the theory has so many new aspects which make 
the theory more interesting. But most importantly our 
results strongly imply that we need a new interpretation 
of the Skyrme theory. 

{\bf Note Added:} One of the referees suggested that 
there might be a strong similarity between the Hopf 
map $S^3 \rightarrow S^2 \rightarrow S^2$ 
discussed by Adam {\it et al.} \cite{adama} and our
result that the baryon number could be decomposed 
to the monopole number and the shell number. 
Although there is no direct relation between this 
work and our result, the Hopf fibering $S^3 
\simeq S^2 \times S^1$ does play the central role
for us to justify the existence of the shell number, 
as we have emphasized in this paper. The details 
of the above results and the questions raised in 
this paper will be discussed in a separate 
publication \cite{cho}.
  
{\bf ACKNOWLEDGEMENT}

~~~The work is supported in part by the National Natural 
Science Foundation of China (Grant 11575254), Chinese 
Academy of Sciences Visiting Professorship for Senior 
International Scientists (Grant 2013T2J0010), National 
Research Foundation of Korea (Grants 2015-R1D1A1A01-057578 
and 2015-R1D1A1A01-059407), and by Konkuk University.


\begin{thebibliography}{99}
\bibitem{sky} T.H.R. Skyrme, Proc. Roy. Soc. (London) {\bf 260}, 127
(1961); {\bf 262}, 237 (1961); Nucl. Phys. {\bf 31}, 556 (1962).
\bibitem{witt} G. Adkins, C. Nappi, and E. Witten,
Nucl. Phys. {\bf B228}, 552 (1983).
\bibitem{rho} A. Jackson and M. Rho, Phys. Rev. Lett. {\bf 51},
751 (1983).
\bibitem{bogo} E.B. Bogomol’ny and V.A. Fateev, Sov. J. Nucl. Phys. 
{\bf 37}, 134 (1983).
\bibitem{prep} See, for example, I. Zahed and G. Brown,
Phys. Rep. {\bf 142}, 1 (1986), and references therein.
\bibitem{man} N.S. Manton, Phys. Lett. {\bf B192}, 177 (1987).
\bibitem{bat} R.A. Battye and P.M. Sutcliffe, Phys. Rev. Lett. {\bf 79}, 
363 (1997).
\bibitem{houg} C.J. Houghton, N.S. Manton, and P.M. Sutcliffe,
Nucl. Phys. {\bf B510},507 (1998).
\bibitem{piet} N.S. Manton and B. Piette, in {\it Proceedings of
the European Congress of Mathematics}, Barcelona (2000), 
edited by C. Casacuberta et al., Progress in Mathematics Vol. 201
(Birkhauser, Basel, 2001). 
\bibitem{sut} R.A. Battye and P.M. Sutcliffe, Phys. Rev. Lett. 
{\bf 86}, 3989 (2001); Rev. Math. Phys. {\bf 14}, 29 (2002).
\bibitem{sky108} D.T.J. Feist, P.H.C. Lau, N.S. Manton, Phys. Rev. 
{\bf D87}, 085034  (2013).
\bibitem{hoyle} F. Hoyle, Astrophys. J. Suppl. {\bf 1}, 121 (1954).
\bibitem{epel} E. Epelbaum, H. Krebs, D. Lee, and U. Meissner,
Phys. Rev. Lett. {\bf 106}, 192501 (2011).
\bibitem{lau} P.H.C. Lau and N.S. Manton, 
Phys. Rev. Lett. {\bf 113}, 232503 (2014).
\bibitem{hal} C.J. Halcrow and N.S. Manton, JHEP {\bf 1501}, 
016 (2015).
\bibitem{adam} C. Adam, J. Sanches-Guillen, and 
A. Wereszczynski, Phys. Lett {\bf B691},105 (2010); 
Phys. Rev. Lett. {\bf 111}, 232501 (2013). 
\bibitem{prl01} Y.M. Cho, Phys. Rev. Lett. {\bf 87}, 252001 (2001).
\bibitem{plb04} Y.M. Cho, Phys. Lett. {\bf B603}, 88 (2004).
\bibitem{ijmpa08} Y.M. Cho, B.S. Park, and P.M. Zhang, Int. J. 
Mod. Phys. {\bf A23}, 267 (2008).
\bibitem{nitta} S.B. Gudnason and M. Nitta, Phys. Rev. {\bf D90}, 
085007 (2014).
\bibitem{plb07} Y.M. Cho, Phys. Lett. {\bf B644}, 208 (2007).
\bibitem{ruo} J. Ruostekoski and J.R. Anglin, Phys. Rev. Lett. {\bf 86}, 
3934 (2001); R.A. Battye, N.R. Cooper, and P.M. Sutcliffe, 
Phys. Rev. Lett. {\bf 88}, 080401 (2002).
\bibitem{pra05}Y.M. Cho, Hyojoong Khim, and Pengming Zhang, 
Phys. Rev. {\bf A72}, 063603 (2005); Y.M. Cho and P.M. Zhang, 
Euro. Phys. J. {\bf B65}, 155 (2008).
\bibitem{adama} C. Adam, B. Muratori, and C. Nash, Phys. Lett.
{\bf B479}, 329 (2000); Phys. Rev. {\bf D61}, 105108 (2000).
\bibitem{cho} Pengming Zhang, Kyoungtae Kimm, J.H. Yoon,
and Y.M. Cho, to be published.
\end{thebibliography}
\end{document}